\documentclass[12pt]{article}            %% LaTeX 2e
\usepackage[dvips]{graphicx}

\newcommand{\be}{\begin{equation}}

\newcommand{\ee}{\end{equation}}

\newcommand{\ba}{\begin{eqnarray}}

\newcommand{\ea}{\end{eqnarray}}

\title{\bf Analyzing laser-plasma interferograms with a
Continuous Wavelet Transform Ridge Extraction technique: the
method}

\author{\bf{ P. \, Tomassini, A.\, Giulietti and  L.A.\, Gizzi} \\
Intense Laser Irradiation Laboratory - IFAM CNR\\
  Area della Ricerca di Pisa\\
  Via G. Moruzzi, 1 56124 Pisa (Italy)\\
   {\it E. mail: tomassini@ifam.pi.cnr.it}\\
\bf{M.\, Galimberti and D.\, Giulietti} \\ Intense Laser
Irradiation Laboratory - IFAM CNR\\
   Dip. di Fisica Universita' di Pisa  and I.N.F.M sez. di Pisa\\
\bf{M.\, Borghesi}\\
 The Queen's University, Belfast (UK)\\
\bf{O.\, Willi}\\
 Blackett Laboratory and Imperial College, London (UK)}

\begin{document}
\maketitle

\begin{abstract}

Laser-plasma interferograms are currently analyzed by extracting
the phase-shift map with FFT techniques (K.A.Nugent, Applied
Optics {\bf 18}, 3101 (1985)). This methodology works well when
interferograms  are only  marginally affected by noise and
reduction of fringe visibility, but  it can fail in producing
accurate phase-shifts maps  when dealing with low-quality images.

 In this paper we will present a
novel procedure  for the phase-shift map computation  which makes
an extensive use of the Ridge Extraction in the Continuous Wavelet
Transform (CWT) framework. The CWT tool is {\it flexible} because
of the wide adaptability of the analyzing basis and it can be very
{\it accurate} because of the intrinsic noise reduction in the
Ridge Extraction.

 A comparative
analysis of the accuracy performances of the new tool and the
FFT-based one shows that the CWT-based tool phase maps are
considerably less noisy and it can better resolve local
inhomogeneties.

\end{abstract}

\section{Introduction}

Interferometric techniques are widely used to characterize the
optical properties of a variety of  media.  An important class of
applications concerns the investigation of the density
distribution of plasmas produced by high intensity laser-matter
interactions.  In recent years various interferometer schemes have
been developed and successfully applied to the characterization of
the wide range of plasma condition which can be achieved in
laser-plasma experiments, from the long-scalelength underdense
plasma generated by laser explosion of a thin foil target to the
steep, denser plasma generated by short pulse interaction with a
solid target. All these schemes make use of a so-called probe beam
which consists of a laser pulse which probes the plasma at a given
time \cite{Benattar} \cite{Willi}.
 The fringe
pattern must be then analysed to obtain the two-dimensional
phase-shift  which contains the physical information on the
plasma.  Then, provided that appropriate symmetry conditions are
satisfied, inversion techniques can be applied to the phase shift
map to obtain the density profile \cite{Abel}.

The traditional way of reading a fringe pattern consisted in
building a grid over the pattern and in measuring, for each
position on the grid, the number of fringe jumps with respect to
the unperturbed fringe structure. This procedure was very simple
and was performed manually. However, the amount of information
which could be extracted in this way was very limited due to the
small number of grid points that can be employed.

In 1982 a novel fringe analysis technique was proposed
\cite{Takeda} in which the phase extraction was carried out using
a procedure based upon Fast Fourier transform (FFT). This
technique allows the information carried by the fringe pattern to
be decoupled by spatial variations of the background intensity as
well as by variations in the fringe visibility, provided that the
scalelength of such perturbations is large compared to the fringe
separation. A few years later this FFT technique was applied for
the first time to laser-produced plasmas \cite{Nugent}. More
recently the technique was extensively applied by our group to the
analysis of long-scalelength underdense laser-plasmas
\cite{Gizzi}\cite{Gizzierr}.  The use of ultra-short probe pulses
has considerably reduced  the fringe-smearing effect due to the
motion of the plasma during the probe pulse. This fact allowed to
investigate short-lived phenomena in the propagation of
ultra-short laser pulses through plasmas \cite{Borghesi}.

The extensive use of the FFT based technique carried out by our
group has shown its effectiveness. In some circumstances however,
reduction of fringe visibility, non uniform illumination, noise
and the presence of local image defects  make the FFT based
technique unstable and the results are not fully satisfactory
because of the presence of unphysical phase jumps. In addition
small scale non-uniformities with low departure from the density
background are unlikely to be detected.

 In this paper we show
that Continuous Wavelet Transforms can also be  applied to the
analysis of interferograms resulting in a  more flexible and
reliable  technique than the FFT based one. To our knowledge, this
is the first time that such an approach is applied to fringe
pattern analysis.

In section 2 we will shortly introduce the Continuous Wavelet
Transform (CWT) and its remarkable properties of good space-scale
analyzer.

In section 3 we introduce our  {\it IACRE}, "Interferogram
Analysis by Continuous wavelet trasform Ridge Extraction"  tool
for interferograms analysis  and we compare its sensitivity to the
FFT-based method.

Section 4 is devoted to conclusions and comments.

\section{The Continuous Wavelet Transform analysis tool }

The Continuous Wavelet Transform is a tool to obtain a
representation of signal $s$
 which is intermediate between the "real time"
description $s = s(t)$ and the "spectral" description $\hat{s} =
\hat{s}(\omega)$, so that it is a very powerful tool to obtain a
time-frequency description of a sequence of data. In this paper
the words "time" and "space", or "frequency" and
"spatial-frequency" will be used indifferently.

 The need of a time-frequency  description of a sequence is much strong
  when the signal represents
a sum of frequency-modulated components (as for each section of an
interferogram image, as shown below). This is because in a  purely
spectral analysis the frequency content of a modulated sinusoid is
generally spread in a large region and no identification of the
signal from its spectral amplitude is allowable.

  The obvious step that
  can be made to overcome the lack of time sensitivity is the
  introduction of a sequence of windows of a given width and
  centered at different times: for each window the FFT of the
  signal is computed and a partial time resolution is obtained.
  These techniques are called "Short-Time Fourier Transform" or
   "Gabor Transform" \cite{Gabor}.
The Gabor Transform is currently used in many context but is not
considered by the signal-processing community a "full analysis
tool".  This is because the number of oscillations of each sinus
in the window depends on the frequency and consequently the
spectral and spatial resolutions should be optimized (by tuning
the window length) only in a narrow band.

From the early  80's, with the introduction of the Wavelet
Transform, a satisfying time-frequency analysis tool is available
\cite{W1} \cite{W2} \cite{W3}.

To introduce the Wavelet Transform, let us first define notations
for the Fourier Transform. For a  signal $s \, \, \epsilon
  \, L^1({\cal R}){\cap} L^2({\cal R})$ the Fourier coefficients,
  that is the scalar product between the signal and the infinitely
  oscillating terms $e_\omega = \exp({-i\, \omega t})$:
  \be
  \hat{s}(\omega) \equiv <e_\omega|s> =\int_{-\infty}^{\infty} dt
  \exp({-i\, \omega t}) s(t)
  \ee
  \noindent form a complete basis of the space to
  which $s$ belongs.

Let us introduce a function $\Psi(t)$ called {\it Mother wavelet}.
Now, instead of decomposing the signal $s$ as a sum of the pure
oscillating terms $e_\omega$ ({\it Fourier Transform}), we build a
decomposition of $s$ {\bf in terms of the base of all the
translated} (by parameter b) {\bf and scaled} (by parameter a)
$\Psi$'s. The base of the Continuous Wavelet Transform ({\it CWT})
is then a two-parameter family of functions
\be
   \label{wav1}
\Psi_{a , b}(t) \equiv {1\over a}\, \Psi\left( (t-b)\over a\right)
\, . \ee

The choice of the Mother Wavelet used to build the analyzing base
is quite free and must be adapted to the actual information that
should be extracted from the signal\cite{W2}.

Once the  base has been built, one can compute the CWT
coefficients as the scalar product of the signal and $\Psi_{a,b}$:
  \ba
  \label{wav2}
  W_{s}(a,b) & \equiv & <\Psi_{a , b}|s> \\
   \nonumber & = &\int_{-\infty}^{\infty} dt
   {1\over a}\,\overline{ \Psi\left( (t-b)\over a\right)} \, s(t), \, \,
  \ea

A particular choice of Mother Wavelet is the Morlet wavelet, and
is largely used in studying signals with strong components of pure
sinus or modulated sinusoids. The Morlet base has the form
\be
\label{Morlet} \Psi(t) = \exp({i\, \omega_0 t}) \, \,
\exp({-(t/\tau)^2})\, , \ee \noindent where the parameters
$\omega_0$ and $\tau$ control the peak frequency and the width of
the wave respectively. The product $\omega_0 \times \tau$ {\it
controls the time and spectral resolution of the Wavelet
decomposition}: a large $\tau$ corresponds to
  a long wave (high spectral resolution and low temporal resolution)
   while a small $\tau$ produces  an "event based" analysis
    (low spectral resolution and high temporal resolution).

We now face the problem of a numerical computation of the
  Wavelet coefficients map $W_s(a,b)$. For a sequence of $N$
   samples $s_i;\, i=1...N$ of $s$, the translation parameter
   $b$ (which controls the central position of the wave envelope)
    can be sampled in a straightforward way:
     $b \rightarrow b_i; \, i=1...N$.
     The scaling parameter $a$
      (which controls the characteristic scale of the wave) may be
sampled as  $\, \, \, \, a_j\, = \,  2^{-j/N_v} \, \, j=1...M \,
,$ (Natural or Log sampling),
  where $N_v$ is the "number of voices per octave" parameter.
  Each $a_j$ is called "voice" and, in the case
   $N_v = 12$, Log  sampling exactly corresponds to the spectral
sampling of musical tones
    in the "tempered scale" introduced by J.S. Bach.

The Log sampling of CWT coefficients in the Morlet basis is very
useful when the spectral content of the signal is the main
information to be extracted, because it provides a good compromise
between spatial and spectral resolution. As the reader can easily
check, the spectral
  resolution at each voice is proportional to the peak frequency of the voice
   $\left(\omega_o \over a\right)$ so that the relative spectral
uncertainty $\Delta f\over f$ is constant along the $a$ axes.

  The real part
of CWT map shows an important feature of CWT with the Morlet base:
${\cal R}(W_s)$ is almost constant, apart from the thin band
centered at the  local signal frequency. Futhermore, the sequence
\be
\label{Ridge1}
  R_s(b) \equiv {\cal R}(W_s)(b,a_R(b))
\ee \noindent where, for each $b^*$, $a_R(b^*)$ is the voice
corresponding to a \underline{local maximum} of the the line-out
of the absolute value CWT map taken at $b=b^*$, well reproduces
the input signal itself. The sequence (ore more generally the
sequences when more complex signals are analyzed) (\ref{Ridge1})
is called {\it the Ridge of CWT map} and represents the subset of
CWT map where most of the "energy" is contained. Presently, the
Ridge detection  of the CWT map plays a rising role in signal
processing \cite{WT4}\cite{IT}, especially in the search of
non-stationary signals with a very low signal-to-noise ratio (see
\cite{IT2} and references therein). This is because the Ridge
sub-map well captures the "true" input signal even in the presence
of a quite strong noise. Futhermore, Ridge extraction in CWT maps
of analytic signals  represents a natural way to detect the local
frequency evolution and, eventually, to easily recover phase
information.

\section{The new CWT-based method.}

\subsection{The FFT-based method for phase-shift estimation}
\begin{figure}
\begin{center}
\includegraphics[angle=0, width=17cm]{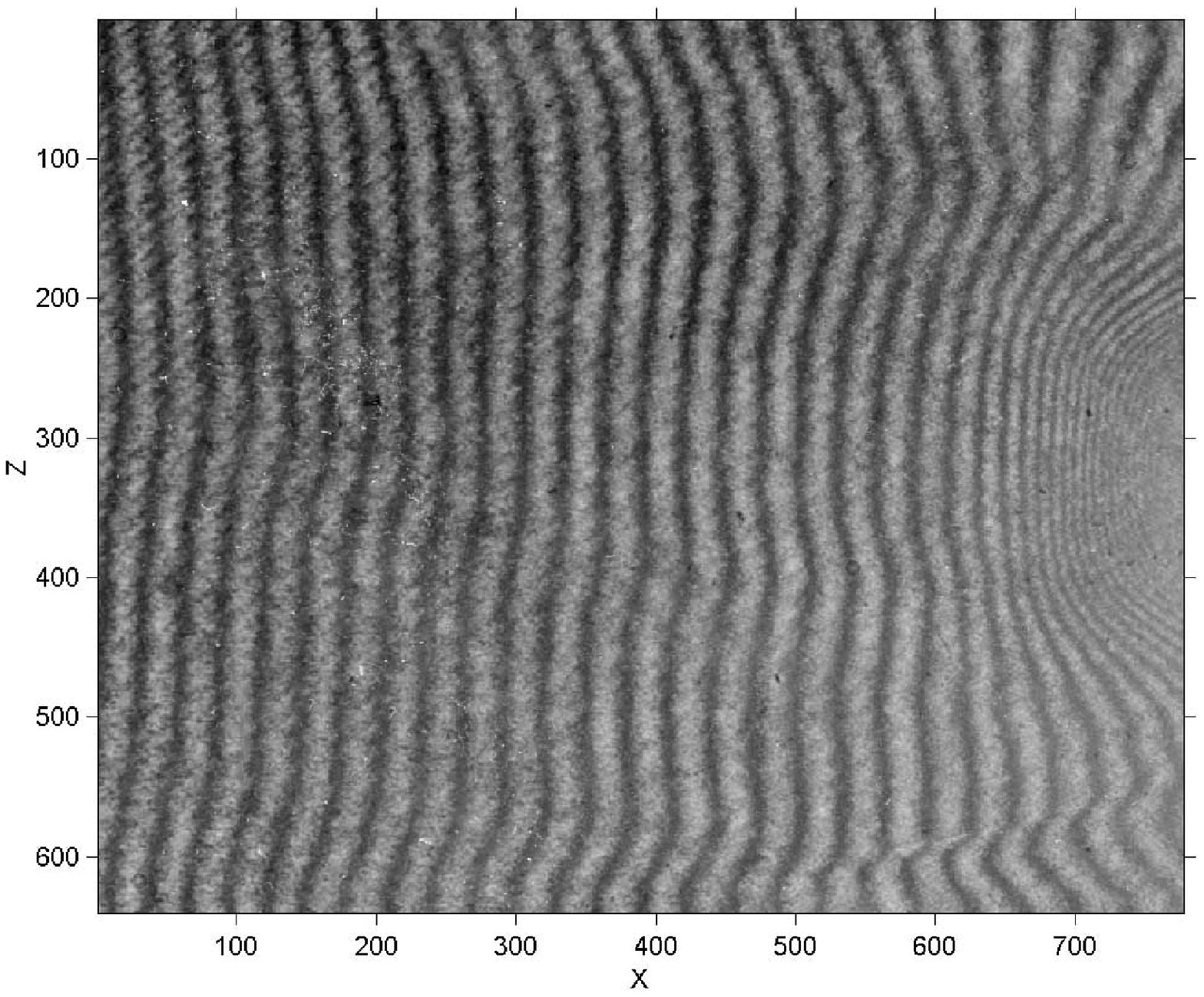}
\caption{}\label{fig:interferogramma}
\end{center}
\end{figure}

\begin{figure}
\begin{center}
\includegraphics[angle=0, width=14cm]{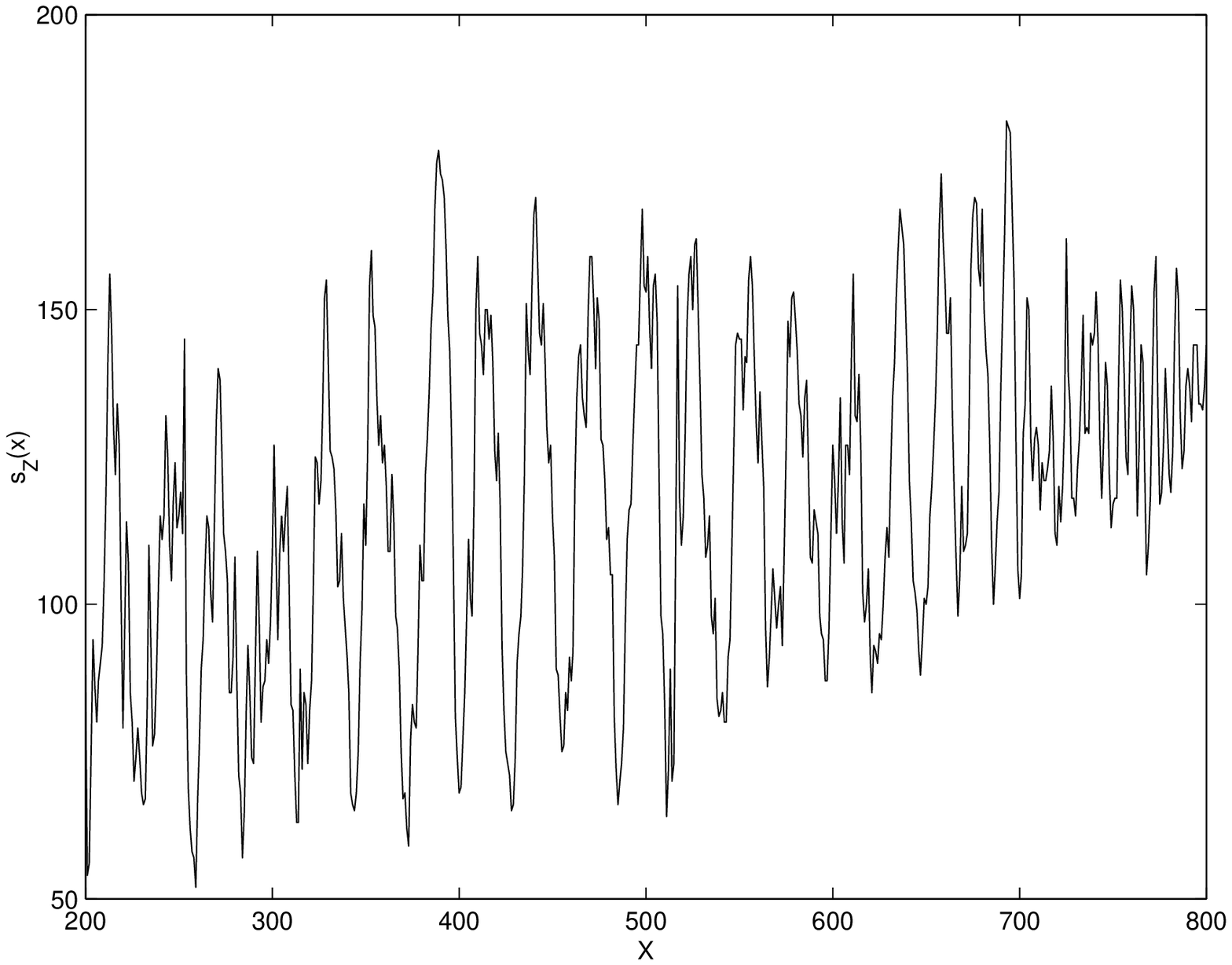}
\caption{}\label{fig:intz}
\end{center}
\end{figure}

The extraction of  phase-shift map, that is the computation for
each pixel of an interferogram  image of the phase-shift with
respect to a unperturbed wave profile, is usually performed with
the help of Fast Fourier Transform (FFT-based method). Consider
for example the interferogram of Fig. \ref{fig:interferogramma} of
a laser-exploded foil target plasma \cite{Gizzi}. Let its
gray-level map be $I(z,x)$ and for each $Z$ build the sequence
$s_Z \equiv I(z =Z,x)$ (that is a horizontal line-out of the
figure). The background fringe pattern would give sequences $s_Z$
very similar to pure oscillating terms plus noise and (possibly) a
slowly varying background. If the departure of such a behaviour is
identified as a local frequency modulation of the oscillating
term, then the phase-shift $\delta\phi(z,x)$ can be easily
computed  as the difference between the perturbed phase at each
$x$ position and the corresponding phase of the background
sequence. Figure \ref{fig:intz} shows a sequence $s_Z$ for $Z =
400$ (the middle of the frame). The behaviour of $s_Z$ can be
identified as a frequency-modulated oscillation with local
frequency $\Omega(x)$ increasing with $x$, plus noise and slowly
rising background. In addition, the amplitude of oscillations
sharply reduces for $x \approx 700$ (this phenomenon is known as
"reduction of fringe visibility", see \cite{Gizzi}).

The FFT-based phase-shift extraction uses FFT for both filtering
the sequence from noise and background (with cuts in the spatial
frequency domain) and extracting the phase by using
straightforward FFT coefficients manipulations \cite{Gizzi}.

\subsection{The {\it IACRE} phase-shift  estimation: an introduction}
To introduce the {\it IACRE} method ("Interferogram Analysis by
Continuous Wavelet Transform Ridge Extraction"), let us observe
that the sequence $s_Z$ (and generally each sequence $I(z = Z,x)$)
has the  structure of a frequency-modulated sinusoid plus some
corrections (noise, slowly varying background). It is  therefore
natural to try to extract the $s_Z$ phase-shifts by using CWT
techniques, with Ridge detection playing a relevant role.

\begin{figure}
\begin{center}
\includegraphics[angle=0, width=16cm]{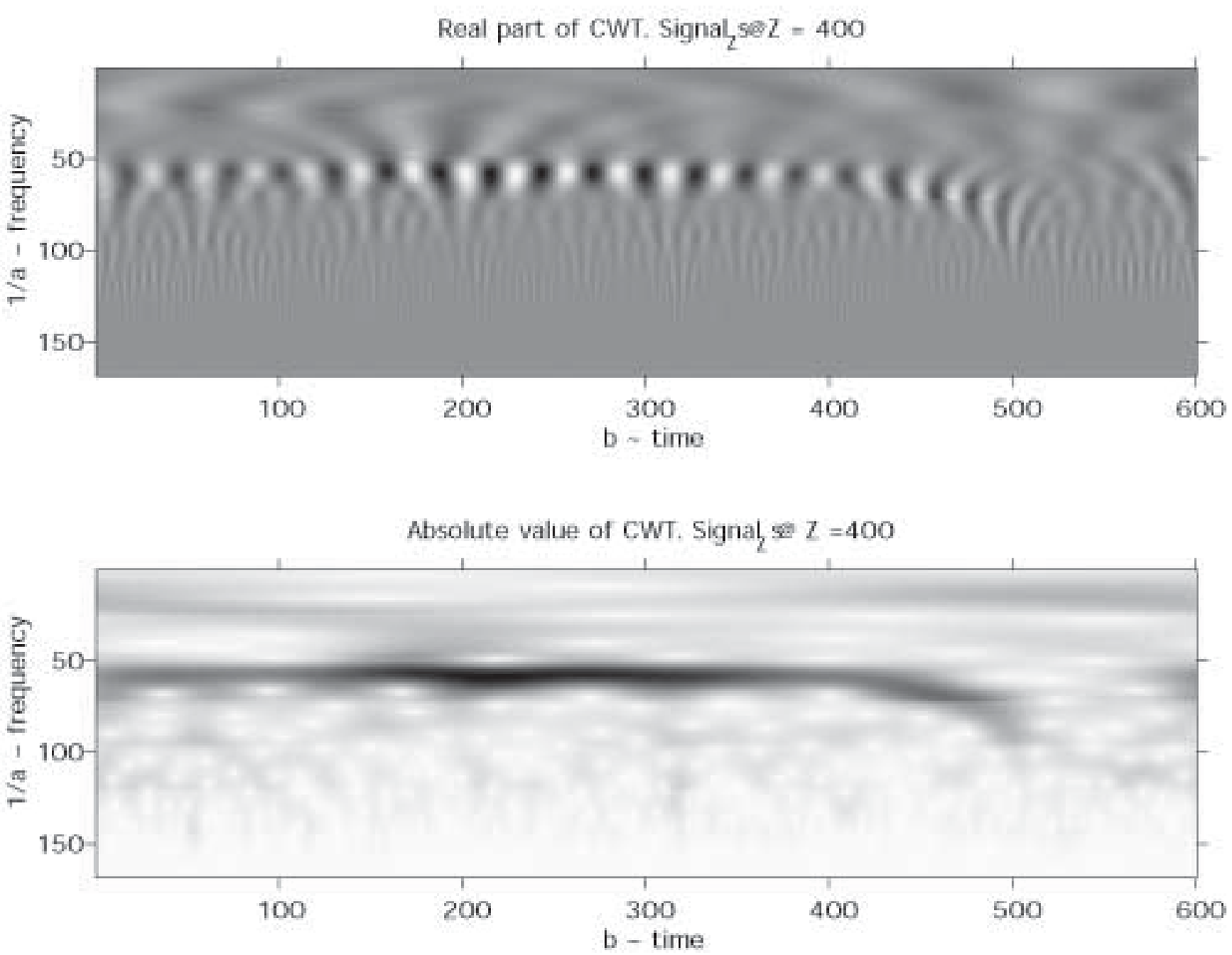}
\caption{} \label{fig:cwtsz}
\end{center}
\end{figure}

\begin{figure}
\begin{center}
\includegraphics[angle=0, width=16cm]{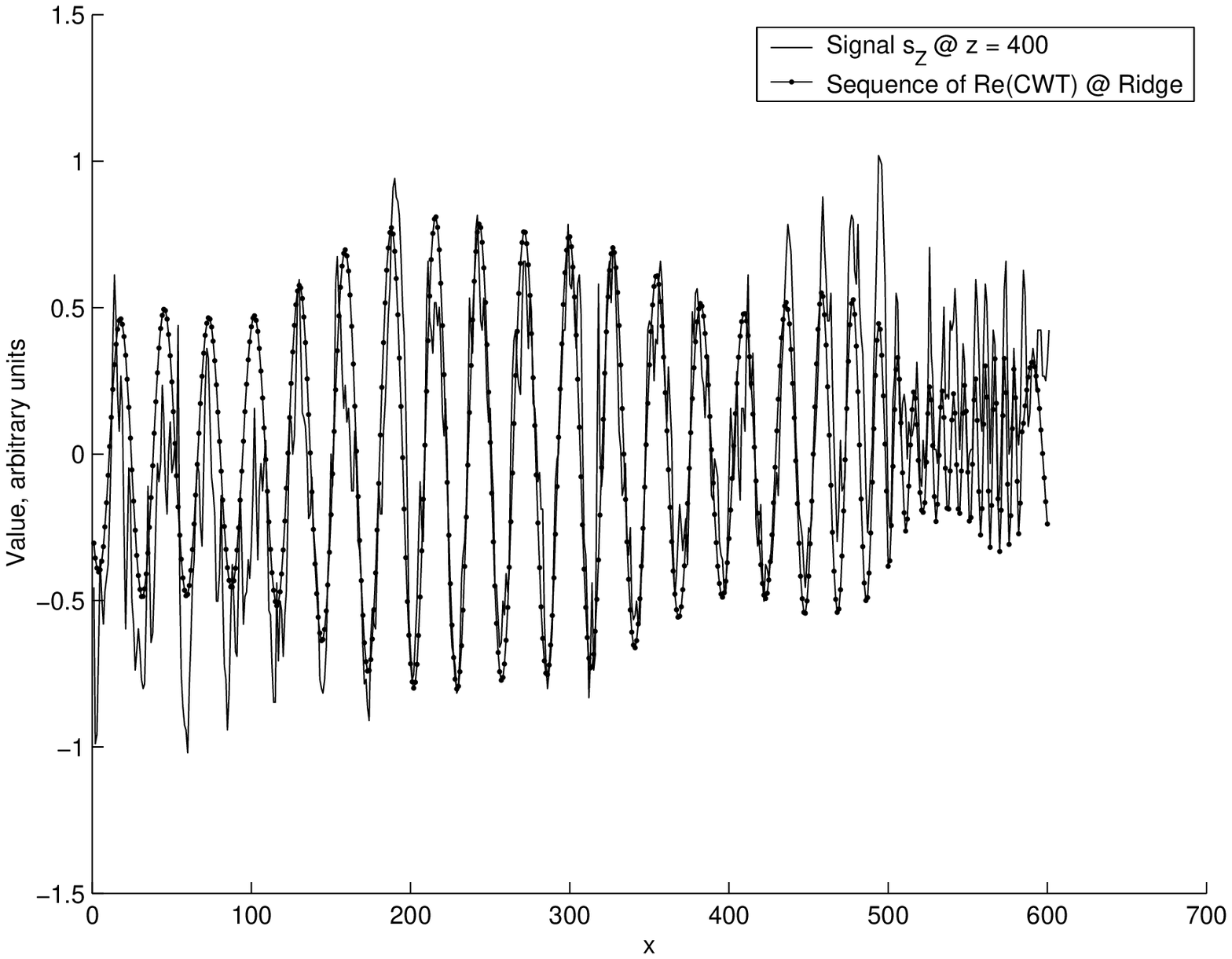}
\caption{}\label{fig:voicesz}
\end{center}
\end{figure}

Consider the CWT map of the sequence $s_Z\, @Z =400$ (see Fig.
\ref{fig:cwtsz}). We can then try to apply the Ridge-extraction
technique to the  CWT map of $s_Z$ to both denoise the sequence
and extract the phase for each pixel position $x$. The Ridge
sequence will be constituted by only the frequency-modulated
components of $s_Z$, so that noise and background will be
automatically discarded. This is the case for $s_Z \, @ Z = 400$,
as it is clear in Fig. \ref{fig:voicesz}. The phase sequence
$\phi_Z(x)$ for the analyzed array $s_Z$ is then simply computed
as the phase of the complex sequence of CWT at the Ridge:
\be
\label{phi1}
  \phi_{s_Z}(x) \equiv phase\left((W_{s_Z})(x,a_R(x)) \right)\, ,
  \ee
  \noindent and the phase-shift $\Delta \phi_{s_Z}(x)$ is obtained
  as
  \be
  \label{phi2}
\Delta \phi_{s_Z}(x) \equiv  \phi_{s_Z}(x)-\phi_0(x) \, , \ee
\noindent where $\phi_0(x) = k_p \, x$ and $k_p$ is the wavenumber
of the not-perturbed fringes.

\subsection{The {\it IACRE} method step-by-step}
  Let us now examine the recipe for the {\it IACRE} algorithm. Let
  $I(z,x)$ be  the gray-level image matrix of dimension $M \times N$.
  The first steps are
  the estimation of the unperturbed fringe wavelength $k_p$ and (eventually)
  image filtering to slightly reduce noise. Next,
  for each $Z \in [1 \,M]$ we consider the sequence $$s_Z(x) \equiv I(z=Z,x)\, \, ,
x \in [1\, N]$$ and:
  \begin{itemize}
  \item{} {\bf Compute the (complex) CWT map $W_Z(a,b)$ with the
Morlet base in the Log sampling}. To do this
  one must choose the number of voices per octave $N_v$. A large
  $N_v$ ($N_v > 12$) should be preferred if fast changes in the local frequency
  $\Omega(x)$ are expected.
  In addition, if we expect that in some
  regions the local frequency $\Omega_Z(x)$ could have abrupt
  changes (local irregularities, structures, edges ...), a higher
  spatial resolution is preferred $(\omega_0 = 2\pi$, $\tau < 1 $),
  while for regular behaviour (like the one of interferogram {\it
  Int 1}) a medium space-frequency resolutions should be used
  ($\tau = 1$).
  \item{} {\bf Detect the (complex) Ridge sequence} $R_Z(x) \equiv
  W_{s_Z}(x,a_R(x))\, .$
  \item{} {\bf Compute the phase of} $R_Z$: $$\phi_Z(x)= phase(R_Z(x))\, .$$
  \item{} {\bf Estimate the phase-shift at $z=Z$ as} $$\delta
\phi(Z,x) \equiv \phi_Z(x)-k_p \, .
  x$$
  \end{itemize}
The result is a phase-shift matrix $\delta\phi(z,x)$ of dimension
$M\times N$. Phase unwrapping algorithms are then applied to the
phase-shift map to eliminate unphysical phase jumps (this is the
case for FFT-based results too).

 \subsection{Comparison between the {\it IACRE} and FFT-based performances}
We now apply the CWT-based and FFT-based methods  to both real and
simulated interferogram images. To start, we apply the {\it IACRE}
method to the whole interferogram of Fig.
\ref{fig:interferogramma}, which is corrupted from noise and shows
strong reduction of fringe visibility and the presence of small
scale periodical structures not related to the plasma properties.

\begin{figure}
\begin{center}
\includegraphics[angle=0, width=19cm]{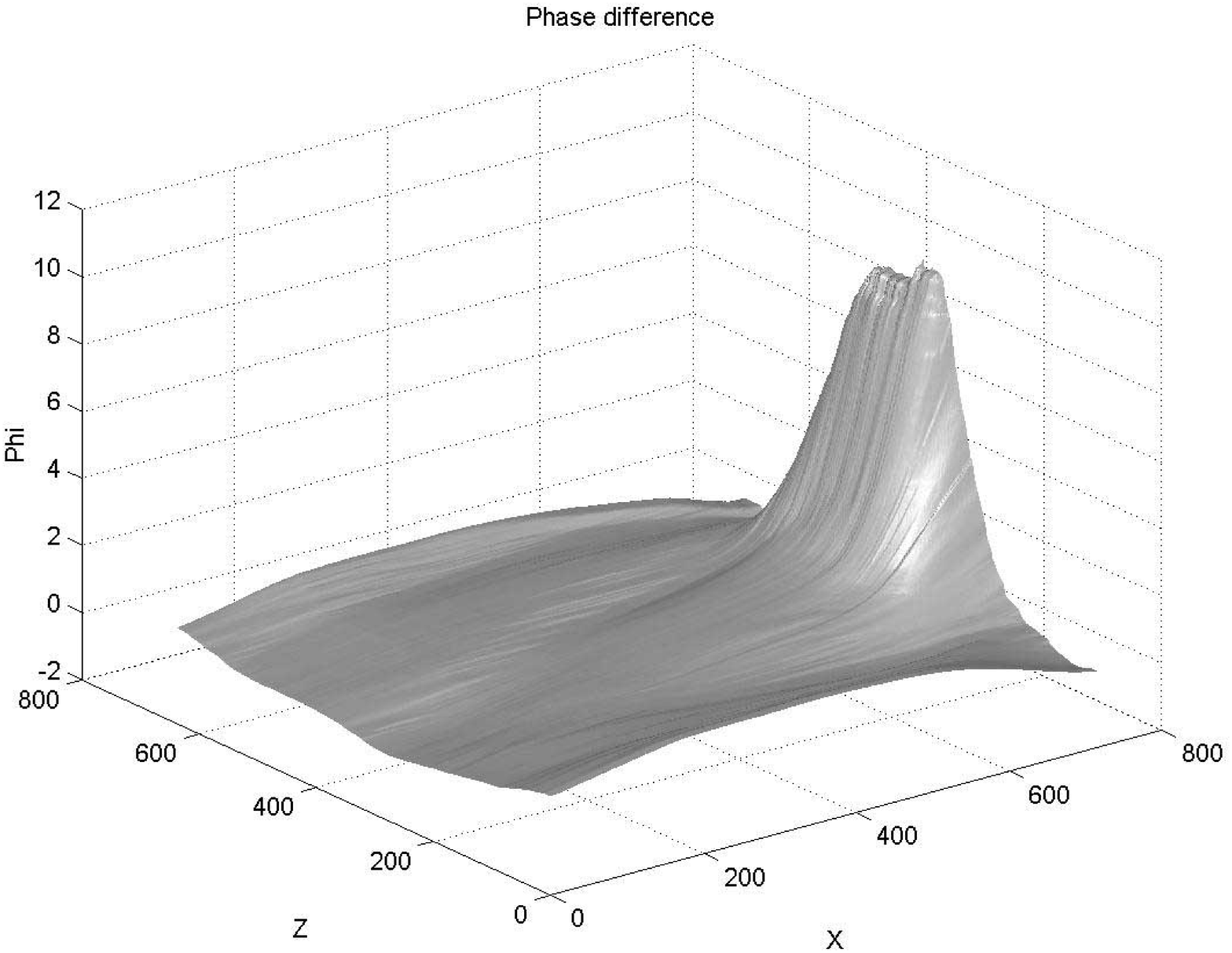}
\caption{} \label{fig:phiCWT1}
\end{center}
\end{figure}

\begin{figure}
\begin{center}
\includegraphics[angle=0, width=19cm]{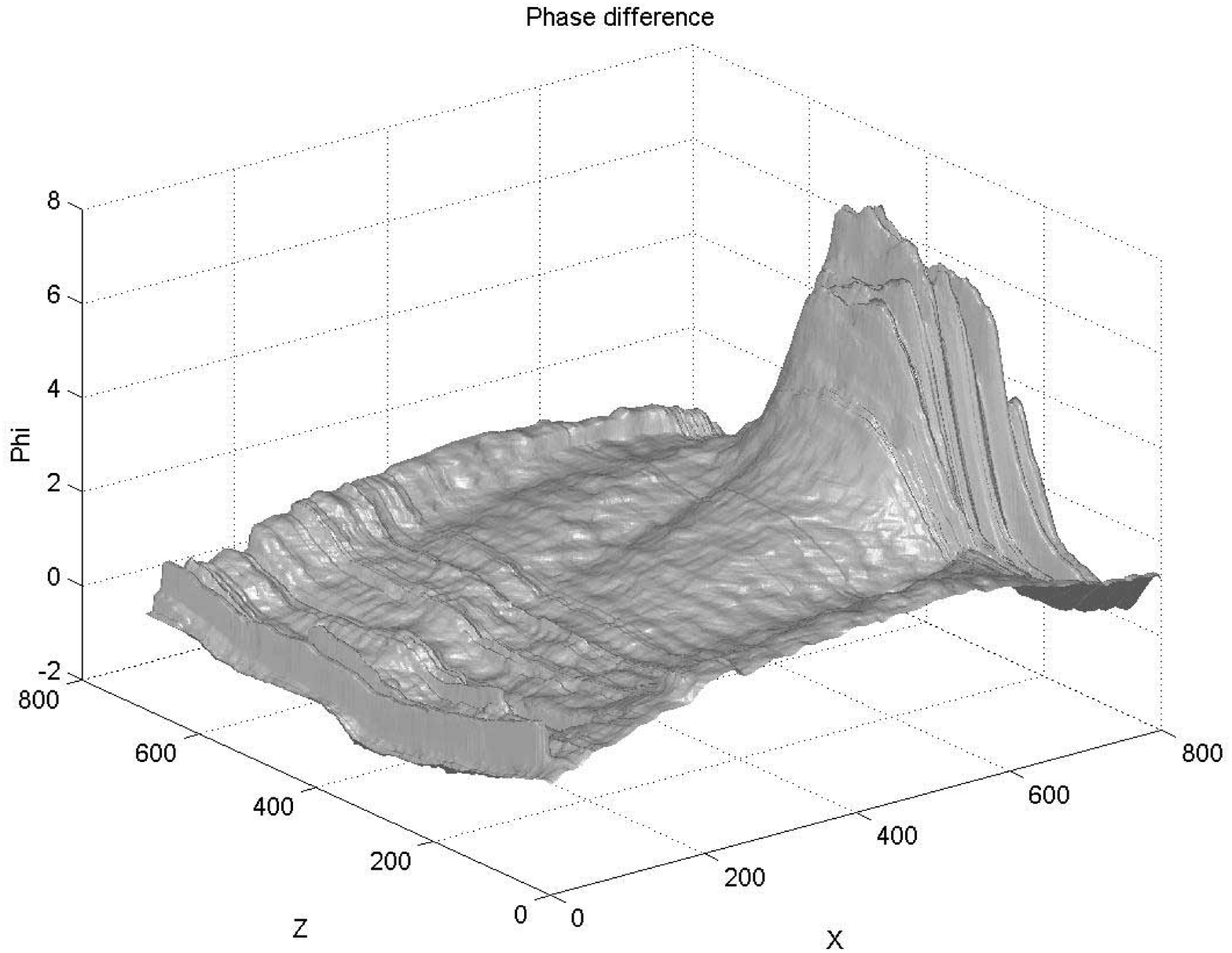}
\caption{}  \label{fig:phiFFT1}
\end{center}
\end{figure}

The {\bf IACRE} output result is obtained with the image { \it
partially filtered from noise} with a Median-Filter of mask size
$3\times 3$ pixels followed by a Wiener-Filter $5 \times 5$ and
using $N_v = 12$ voices per octave (low $N_v$).  The phase shift
map is shown in Fig. \ref{fig:phiCWT1}, which should be compared
with the FFT-method phase-shift  of Fig. \ref{fig:phiFFT1}
obtained with the same filtered image. As it is clear from figures
Figg. \ref{fig:phiCWT1} and \ref{fig:phiFFT1}, while the CWT
output seems to be accurate, FFT output is noisy and not free from
unphysical phase jumps near the target, where a strong reduction
of fringe visibility is present.

The higher accuracy of the {\it IACRE} method with respect to the
FFT-based
 one is a very important characteristics of our new procedure. It
enables an accurate search of small non-uniformity of the
phase-shift map which are important to detect the growth of plasma
instabilities as filamentation and self-focusing.

To better check this point, we numerically build-up one
interferogram in which we simulate the phase shift produced by a
slowly-varying background plus some small scale filaments. Noise
and reduction of fringe visibility are finally added to the
interferometric image to better match the real interferograms
characteristics.

\begin{figure}
\begin{center}
\includegraphics[angle=0, width=16cm]{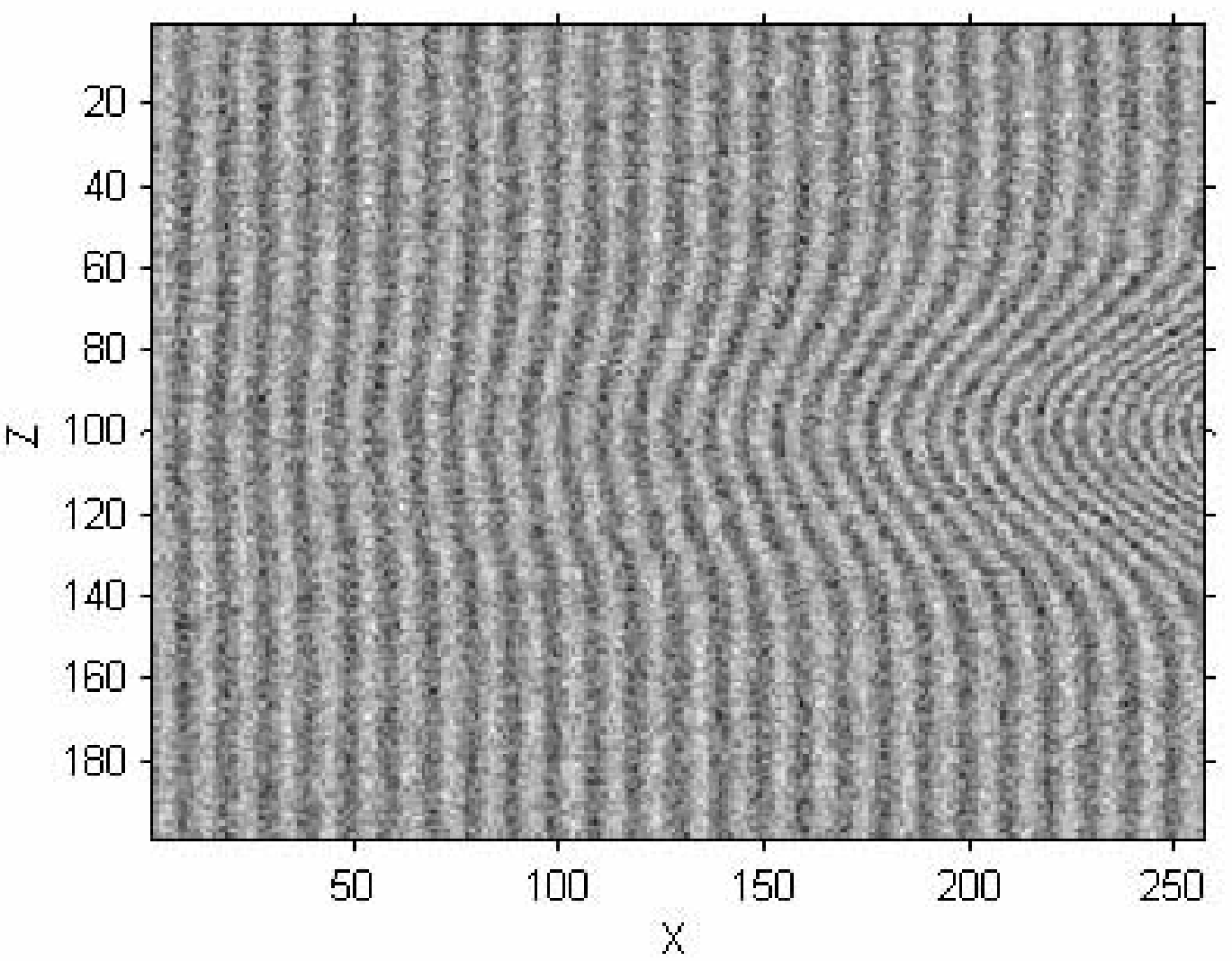}
\caption{}\label{fig:simulato1}
\end{center}
\end{figure}

The  interferogram of Fig. \ref{fig:simulato1}  simulates a plasma
with a  background  of maximum electronic density $(n_e/n_c)_{Max}
= 0.1$ with a Gaussian profile in the radial direction (with
radius $75 \mu m$) which is exponentially decreasing in the $x$
direction.  Three filaments are then added in different positions,
each one with Gaussian density profile: $${\delta n(x,y,z)\over
n_c} = \alpha \exp(-(z^2+y^2)/r^2)$$
 with
maximum density perturbation and radius ($\alpha = 0.005, r= 10
\mu m$),
 ($\alpha = 0.005, r= 8 \mu m$)
and ($\alpha = 0.005, r= 6 \mu m$), respectively. Since the
electronic density is everywhere much lower than the critical
density, the linearity of the phase map with respect to the
density is respected. We can then compute the perturbation of the
phase-shift map in $2 \pi $ units $(\phi_{2\pi} \equiv
\phi/(2\pi)$) with respect to the background as
 $$\delta(\Delta \phi_{2\pi}) = -{1\over 2 \lambda_p} \int{ {\delta n\over  n_{c} } dy} \, ,$$
whose maximum value is
 \be
    \label{maxdeltaphi}
  \delta(\Delta \phi_{2\pi})_{Max} = {1\over 2} \sqrt{\pi} \alpha { r\over \lambda_p}\, ,
  \ee
that is $\delta(\Delta \phi_{2\pi})_{Max}= 0.18 $, $\delta(\Delta
\phi_{2\pi})_{Max}= 0.14  $ and $\delta(\Delta \phi_{2\pi})_{Max}=
0.11,$ respectively. To detect these structures, the noise level
of the phase-shift map should be a fraction of $\delta(\Delta
\phi_{2\pi})_{Max}$. If $\sigma(x)$ is the standard deviation of
the noise of each sequence of $\Delta \phi_{2\pi}(z,x)$ at $x$
fixed, we could detect  these structures if their amplitudes are
for instance at "two sigma" with respect to the noise, that is
$\sigma(x) < 0.09$, $\sigma(x) < 0.07$ and $\sigma_x < 0.05$,
respectively. The standard deviation  $\sigma(x)$ (or one fraction
of $\sigma(x)$) could be then be considered as a rough estimation
of the "detectable threshold" in the phase-shift map".

\begin{figure}
\begin{center}
\includegraphics[angle=0, width=18cm]{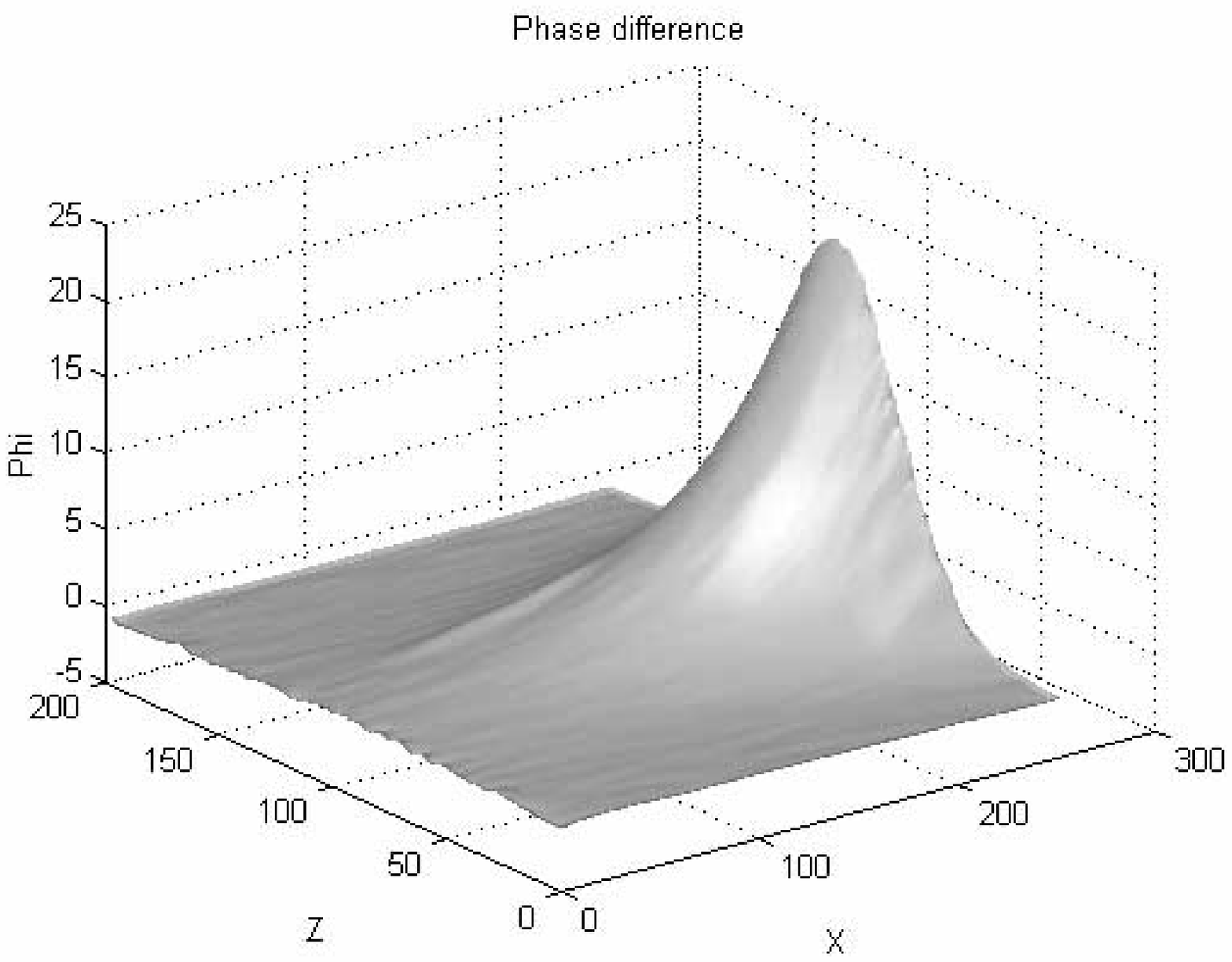}
\caption{} \label{fig:Phsimulato1CWT}
\end{center}
\end{figure}

\begin{figure}
\begin{center}
\includegraphics[angle=0, width=18cm]{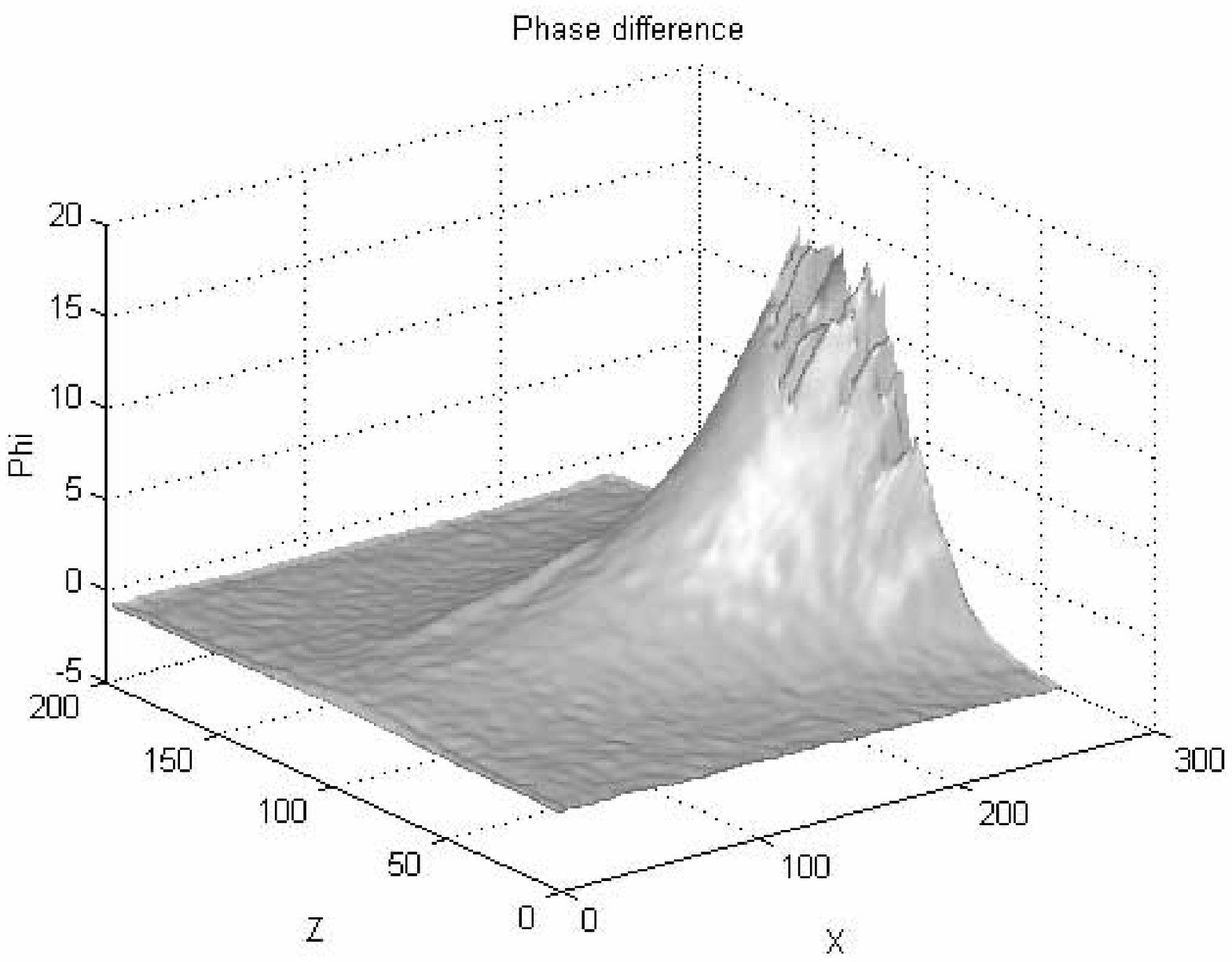}
\caption{}\label{fig:Phsimulato1FFT}
\end{center}
\end{figure}

\begin{figure}
\begin{center}
\includegraphics[angle=0, width=18cm]{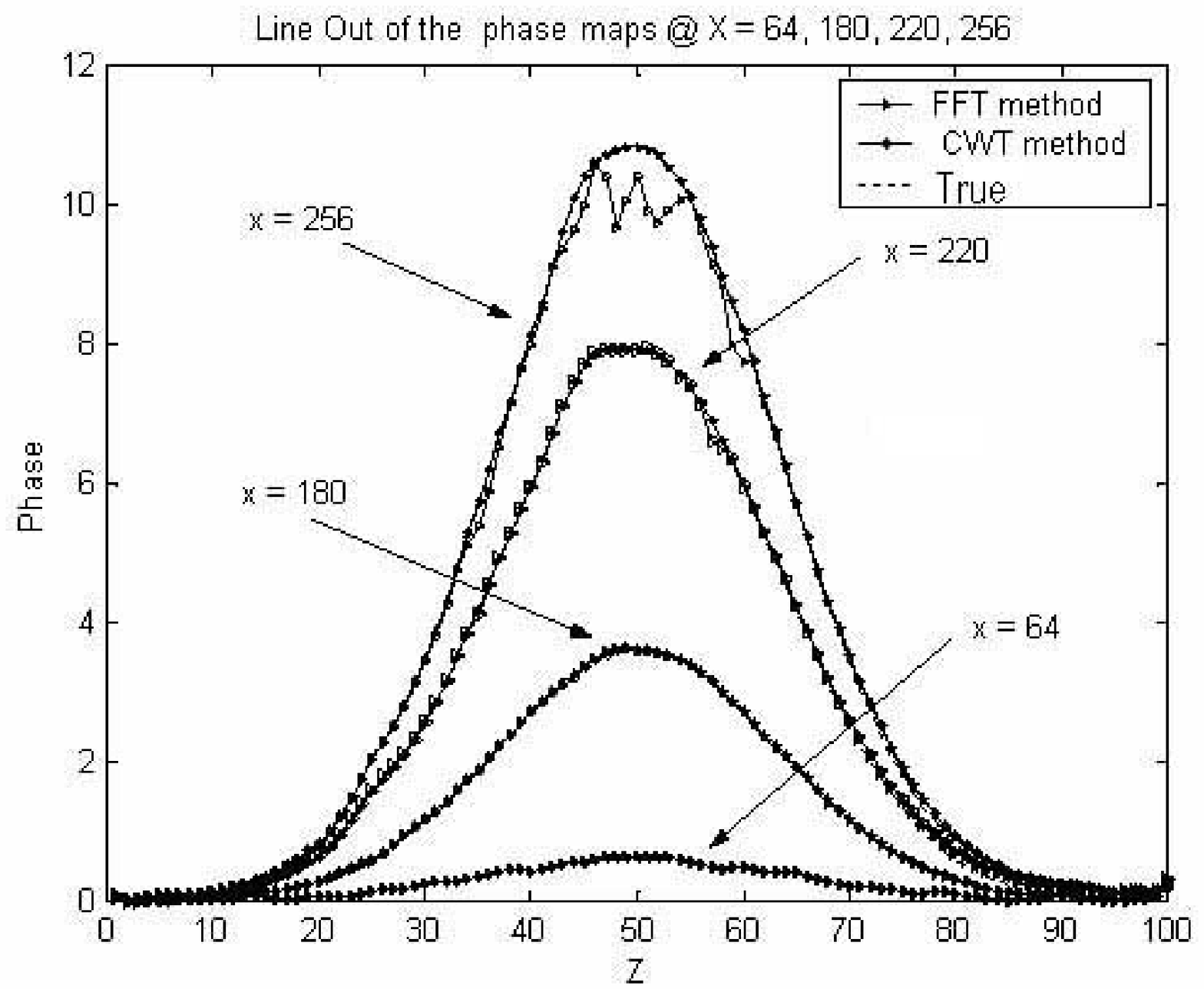}
\caption{}  \label{fig:lineoutsim1}
\end{center}
\end{figure}

To estimate the accuracy of the phase-shift maps obtained with the
{\it IACRE} and FFT-based methods, we compute the phase-shift maps
with the two methods (see Figg. \ref{fig:Phsimulato1CWT} and
\ref{fig:Phsimulato1FFT}) and we compare them with the known
"true" phase map. We start the analysis by comparing some line-out
of the two phase maps with the known simulate map. In  Fig.
\ref{fig:lineoutsim1} it is clear that the accuracy in the two
phase methods is comparable in regions of the interferogram with
low phase-shift, while for large phase-shifts the FFT-based output
clearly fail in producing an accurate phase map.

 Denoting with $\Delta \phi_{2\pi}^{CWT}$ and
$\Delta \phi_{2\pi}^{FFT}$ the phase-shifts maps obtained with the
two methods and with $\Delta \phi_{2\pi}^{Thrue}$ the simulated
phase map, we estimate the error map  as the differences:
 \ba
 \label{errorsim1}
 {\cal E}_{CWT}(z,x) &\equiv&  \Delta \phi_{2\pi}^{CWT}(z,x)-\Delta
 \phi_{2\pi}^{Thrue}(z,x)\, , \nonumber \\
 {\cal E}_{FFT}(z,x) &\equiv&  \Delta \phi_{2\pi}^{FFT}(z,x)-\Delta
 \phi_{2\pi}^{Thrue}(z,x)\, , \nonumber
 \ea
\noindent so that the sequences of the standard deviations of the
noise in the phase map can be estimated as
 \ba \label{sigmanoise}
\sigma_{FFT}(x) &=& std({\cal E}_{FFT}(z,x))\, , \nonumber\\
\sigma_{CWT}(x) &=& std({\cal E}_{CWT}(z,x))\, , \ea \noindent
where $std(f(z))$ is the standard deviation of a sequence $f(z)$.

\begin{figure}
\begin{center}
\includegraphics[angle=0, width=16cm]{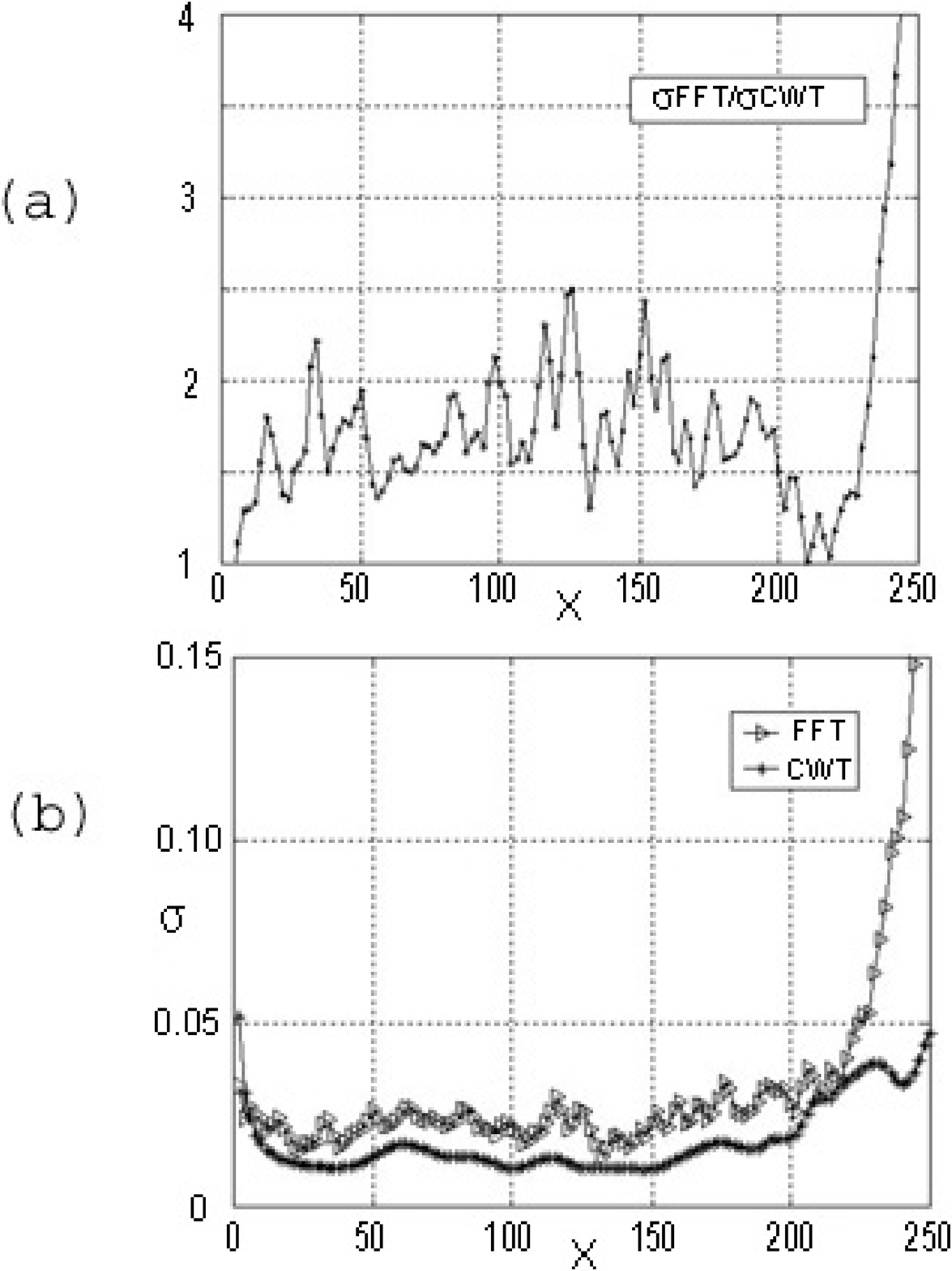}
\caption{}\label{fig:sigmasimulato1}
\end{center}
\end{figure}

In Fig. \ref{fig:sigmasimulato1} is shown the behaviour of the
error in both the {\it IACRE} and FFT-based maps ({\bf (a)}) while
in {\bf b)} the ratio $R(x)$ between $\sigma_{FFT}(x)$ and
$\sigma_{CWT}(x)$ sequences is reported. The analysis of these
figures confirms the claim that in small phase-shifts regions the
{\it IACRE} method exhibits a slightly higher precision than the
FFT-based one (the $\sigma_{FFT}/\sigma_{CWT}$ sequence is about
$2$), while in large phase-shift regions the sensibility of the
{\it IACRE} method is much higher than the one of the FFT-based
one. For example,  assuming the sequence $\sigma(x)$ as an
estimation of the phase-shift sensibility, since for gaussian
density profiles $\delta(\Delta \phi_{2\pi})_{Max}$ is
proportional to the structure radius $r$ and the maximum density
perturbation $\alpha$ (see \ref{maxdeltaphi}), the sequence $R =
\sigma_{FFT}/\sigma_{CWT}$ could be interpreted as a rough
estimation of the ratio between the {\bf minimum} product $\alpha
\, r$ detectable with the FFT-based  and {\it IACRE} techniques:
$$R \equiv {\sigma_{FFT}\over \sigma_{CWT}} \sim {(\alpha \,
r))^{Min}_{FFT}\over (\alpha\, r))^{Min}_{CWT}} \,. $$

Futhermore, since for the {\it IACRE} method the $\sigma_{CWT}$
sequence is everywhere below the value $0.03$ (see the "detection
thresholds" reported above), we are confident  that all the three
filaments could be detected. This is not the case for the
FFT-based method output because in the region $X\in [220, 250]$
the $\sigma_{FFT}$ sequence is in the range $0.05-0.15$, which is
over the minimum of the detection thresholds.

We conclude the analysis of the interferogram of Fig. 7  by
checking the behaviour of the phase maps when an algorithm for the
automatic extraction of small scale perturbations is applied to
the phase-shift maps. The algorithm utilized is very simple and
consists of two main steps:
\begin{itemize}
\item{} The decomposition of the map $\Delta\phi_{2\pi}$ in a 'Large
scale' component (the background) $\bar{\Delta\phi}_{2\pi}$ and a
'Small scale' component  $\delta(\Delta\phi_{2\pi})$ (the
structures $+$ noise) by using a Smoothing B-spline fitting for
each line-out of the phase map at $x$ fixed.
\item{} The filtering of the small scale component
$\delta(\Delta\phi_{2\pi})$ with a "two sigma" cutoff. As
explained before, provided that structures in the
$\delta(\Delta\phi_{2\pi})$ give a negligible contribution in the
Root-Mean-Square of the map, the standard deviations
$\sigma_{FFT}(x)$ and $\sigma_{FFT}(x)$ can be computed as
\ba\label{sigmanoise2} \sigma_{FFT}(x) &=&
std(\delta(\Delta\phi^{FFT}_{2\pi})(z,x))\, , \nonumber\\
\sigma_{CWT}(x) &=& std(\delta(\Delta\phi^{CWT}_{2\pi})(z,x))\, ,
\ea
\end{itemize}

\begin{figure}
\begin{center}
\includegraphics[angle=0, width=12cm]{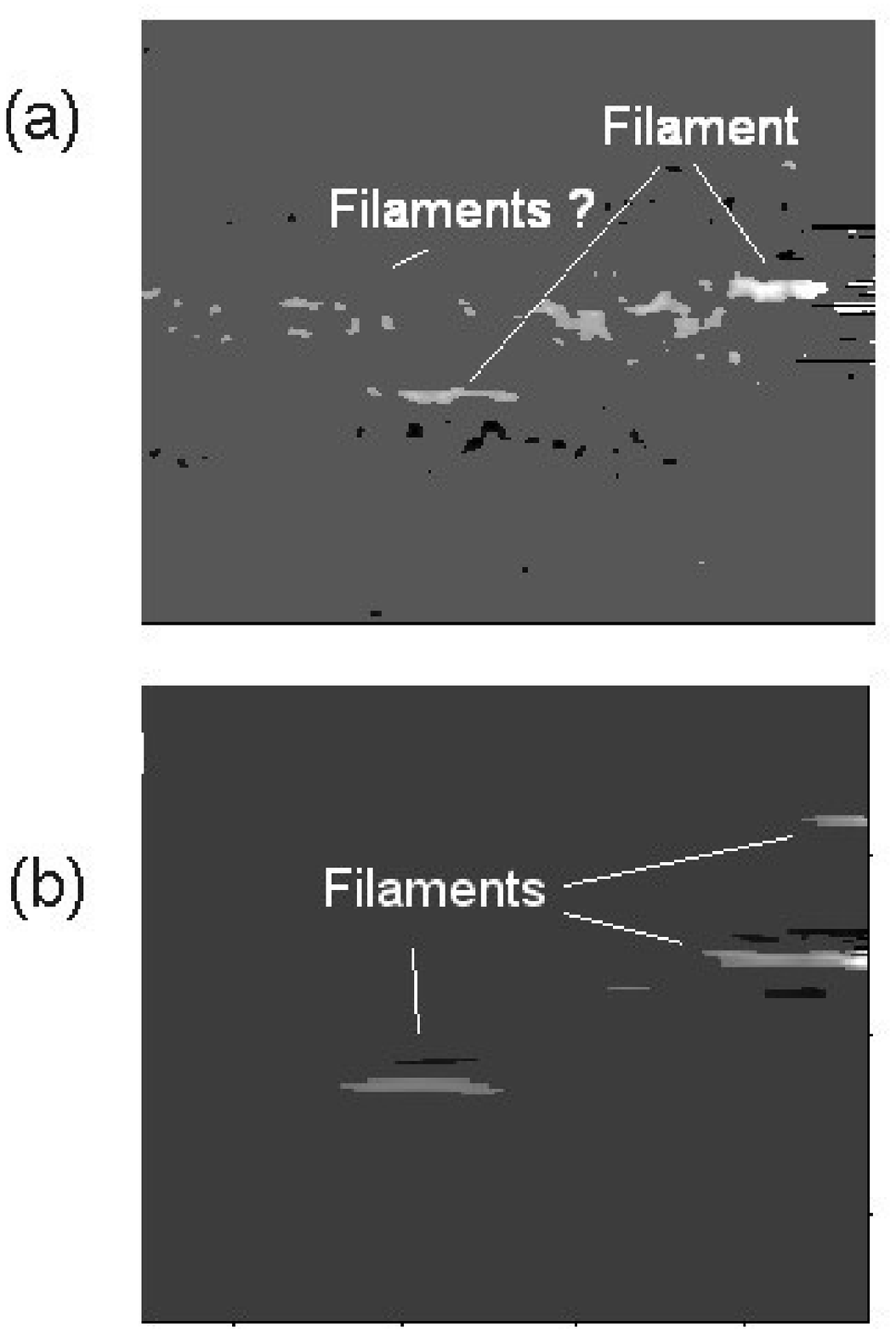}
\caption{}\label{fig:struct1}
\end{center}
\end{figure}

In Fig. \ref{fig:struct1} the filtered at "two sigma" 'Small
scale' phase maps obtained with the two methods are reported. As
expected, the filtered map of the {\it IACRE} method clearly shows
the presence of the three filaments, while in the FFT-based map
some regions of the map with strong presence of noise could be
interpreted as {\it false} small scale structures so no clear
filaments detection is possible.

\section{Conclusions}

 With the help of one real and one simulated interferograms
we showed that the {\it IACRE} method is more {\it accurate} and
{\it robust} than the FFT-based one. For the simulated
interferogram the smallest detectable phase-shift perturbation
(with respect to the background) obtained with the {\it IACRE}
method is in the mean $0.5$ times the one obtained with the
FFT-based one, with possible further decrease in higher density
regions. In addition the outputs of {\it IACRE} are free from
unphysical phase jumps both in the real and the simulated
interferograms, while FFT-based map is in both  cases  affected by
a large region near the target where phase jumps cannot be removed
by conventional unwrapping procedures. The higher {\it robustness}
and {\it sensibility}  of the {\it IACRE} method can be addressed
both to the wide adaptability of the CWT tool to the actual image
and the intrinsic strong noise suppression in the Ridge Extraction
procedure.

\section*{Acknowledgments}
The authors wish to acknowledge support from the italian
M.U.R.S.T. (Project: "Metodologie e diagnostiche per materiali e
ambiente"). One of us (PT) would also thank Guido Buresti
(University of Pisa) and Elena Cuoco (I.N.F.N, section of
Firenze/Urbino) for useful discussions on Continuous Wavelet
Transform.

\newpage
\section*{Figures Caption}

\vspace{1cm}

Fig. $1$ A sample interferogram  of a plasma produced by laser
explosion of a $0.5 \mu m$ thick, $400 \mu m$ diameter Aluminium
dot coated onto a $0.1 \mu m$ plastic stripe support. The
interferogram was taken, perpendicularly to the strip surface,
$3.0 ns$ after the peak of the plasma forming pulses using a
modified Nomarski interferometer. The intensity on target was $8.5
\times 10^{13} W/cm^2$. The probe pulse-length was $100 ps$ and
the probe wavelength was $0.53 \mu m$. For details on the
experimental set-up see \cite{Gizzi}.

\vspace{1cm}

Fig. $2$ Line-out of the fringe intensity (sequence $s_Z$) of the
interferogram of Fig. 1 at $Z = 400$.

\vspace{1cm}

Fig. $3$ The real part and absolute value of  the CWT maps of the
signal $s_Z\, @\, Z = 400$ (interferogram of Fig. 1).

\vspace{1cm}

Fig. $4$  The sequence of real part of the Ridge sequence of
signal $s_Z @ Z = 400$ (interferogram of Fig.1).

\vspace{1cm}

Fig. $5$  The phase-shift map (in $2\pi$ units) obtained from the
interferogram of Fig. 1 after suitable filtering. {\bf IACRE}
method.

\vspace{1cm}

Fig. $6$ The phase-shift map (in $2\pi$ units) obtained from the
interferogram of Fig. 1 after suitable filtering. {\bf FFT-based}
method.

\vspace{1cm}

Fig. $7$ A simulated interferogram  of a plasma containing three
small filaments.

\vspace{1cm}

Fig. $8$  Phase-shift map of the simulated interferogram  of Fig.
7. {\bf IACRE} method.

\vspace{1cm}

Fig. $9$ Phase-shift map of the  simulated interferogram of Fig.
7. {\bf FFT-based} method .

\vspace{1cm}

Fig. $10$ Line-out of the phase-shift maps  of the simulated
interferogram of Fig. 7. The {\bf IACRE} and {\bf FFT-based}
methods outputs are compared with the 'true' simulated map.

\vspace{1cm}

Fig. $11$ {\bf (a)} Standard deviations of the error in the
phase-shift map of interferogram of Fig. 7 computed via IACRE and
FFT-based methods. {\bf (b)} Ratio between the standard deviations
of the error in the phase-shift map of interferogram of Fig. 7
computed via FFT-based and IACRE   methods. For density
perturbations $\delta n$  with gaussian density profile in the $z$
direction of amplitude $\alpha$ and radius $r$, the sequence
$\sigma_{FFT}/\sigma_{CWT}$ can also be interpreted as the ratio
between the {\bf minimum} product $\alpha\,   r$ detectable with
the FFT-based and IACRE methods: $\sigma_{FFT}/\sigma_{CWT} \sim
(\alpha\, r)^{Min}_{FFT}/(\alpha\, r)^{Min}_{CWT} $.

\vspace{1cm}

Fig. $12$ Filtered map at "two sigma" of the 'Small scale'
component of the phase-shift map of interferogram of Fig. 7. {\bf
(a)} FFT-based  method: two filaments could be detected but other
unreal structures survive to the "two sigma" filter. {\bf (b)}
{\it IACRE} method: three filaments are clearly detected.

\newpage

\newpage

\newpage

\newpage

\newpage

\newpage

\newpage

\newpage

\newpage

\newpage
\small{.}

\newpage
\small{.}

\newpage

\newpage


\begin{thebibliography}{99}

\bibitem{Benattar}
{R.Benattar, C.Popovics, R.Siegel, {\it Polarized light
interferometer for laser fusion studies}, Rev.Sci.Instrum.
{\bf50}, 1583 (1979)}
\bibitem{Willi}
{O.Willi, {\it Diagnostics and experimental methods of laser
produced plasmas}, in Laser-Plasma Interaction 4, Proceedings of
XXXV Scottish Universities Summer School in Physics, St.Andrews,
SUSSP Publications, University of Edinburg, 1988}
 \bibitem{Abel}
{P. Tomassini and A. Giulietti, {\it A generalization of Abel
Inversion to non axisymmetric density distribution}, accepted for
pub. on Opt. Comm. }
\bibitem{Takeda}
{M.Takeda, H.Ina, S.Kobayashi, {\it Fourier-transform method of
fringe-pattern analysis for computer-based topography and
interferometry},  J.Opt.Soc.Am. {\bf 72}, 156 (1982).}
\bibitem{Nugent}
{K.A.Nugent, {\it Interferogram analysis using an accurate fully
automatic algorithm}, Applied Optics {\bf18}, 3101 (1985)}
\bibitem{Gizzi}
{L.A.Gizzi,D.Giulietti, A.Giulietti, T.Afshar-Rad, V.Biancalana,
P.Chessa, E.Schifano, S.M.Viana, O.Willi, {\it Characterisation of
Laser Plasmas for Interaction Studies}, Phys.Rev. E, {\bf49}, 5628
(1994) }
\bibitem{Gizzierr}
{L.A.Gizzi,D.Giulietti, A.Giulietti, T.Afshar-Rad, V.Biancalana,
P.Chessa, E.Schifano, S.M.Viana, O.Willi, {\it Characterisation of
Laser Plasmas for Interaction Studies. Erratum},  Phys.Rev. E,
{\bf50}, 4266 (1994) }
\bibitem{Borghesi}
{M.Borghesi, A.Giulietti, D.Giulietti, L.A.Gizzi, A.Macchi,
O.Willi, {\it Characterization of laser plasmas for interaction
studies: progress in time-resolved density mapping}, Phys.Rev. E,
{\bf54}, 6768 (1996) }
\bibitem{Gabor}
{D. Gabor; {\it Theory of Communication}, J. Inst. Electr. Eng.,
London, 93 (III), pp 429-457}
\bibitem{W1}
{J. Morlet, G. Arens, I. Fourgeau and D. Giard; {\it Wave
propagation and sampling theory}, Geophysics, {\bf 47}, pp.
203-236}
\bibitem{W2}
{M. Holschneider; {\it Wavelet: An analysis tool}, Clarendon Press
-Oxford (1995)}
\bibitem{W3}
{I. Daubechies; {\it Ten lectures on Wavelets}, Soc. for Ind. and
Applied Mathematics, Philadelphia (1992)}
\bibitem{WT4}
{ R. Carmona, W.L. Hwang and B. Torresani; {\it Characterization
of Signals by the Ridges of their Wavelet Transform}. paper; IEEE
Trans. Signal Processing {\bf 45}, vol. 10, p. 2586.}
\bibitem{IT}
{B. Torresani;{\it  Time Frequency and Time Scale Analysis},
abstract in Signal Processing for Multimedia, J. Byrnes Ed. (1999)
p. 37-52. }
\bibitem{IT2}
{J.M. Innocent and B. Torresani; {\it A Multiresolution Strategy
for Detection Gravitational Waves Generated by Binary
Coalescence}, Submitted to Phys. Rev. D}

\end{thebibliography}
\end{document}